\documentclass[twocolumn,showpacs,preprintnumbers,amsmath,amssymb]{revtex4}

\usepackage{graphicx}
\usepackage{dcolumn}
\usepackage{bm}

\begin{document}

\preprint{APS/123-QED}

\title{Beyond Gibbs-Boltzmann-Shannon: \\  General Entropies -- The Gibbs-Lorentzian Example}

\author{R. A. Treumann}
 \altaffiliation[Also at ]{International Space Science Institute, Bern, Switzerland}

%
\affiliation{Department of Geophysics and Environmental Sciences, Munich University, Munich, Germany
}%
\email{treumann@geophysik.uni-muenchen.de}
\author{W. Baumjohann}
\affiliation{Space Research Institute, Austrian Academy of Sciences, Graz, Austria
}%
\email{
Wolfgang.Baumjohann@oeaw.ac.at}
\date{\today}

\begin{abstract}
We propose a generalisation of Gibbs' statistical mechanics into the domain of non-negligible phase space correlations.  Derived are the  probability distribution and entropy as a generalised ensemble average, replacing Gibbs-Boltzmann-Shannon's entropy definition enabling construction of new forms of statistical mechanics.  The general entropy may also be of importance in information theory and data analysis. Application to  generalised Lorentzian phase space elements yields the Gibbs-Lorentzian power law probability distribution and statistical mechanics. Details can be found in arXiv:1406.6639. 
\end{abstract}

\pacs{94.05.Lk, 94.20.wj, 05.30.Ch}
\maketitle

\section{Generalisation of Gibbs' Statistical Mechanics}
In Gibbs' Statistical Mechanics \citep{landau1994,huang1987} the probability $w_i(\epsilon_i)\propto\exp(-\epsilon_i/T)$ of finding a particle in energy state $\epsilon_i$ at constant temperature $T$ (in energy units) is obtained from considering the infinitesimal change of the phase space volume $\Gamma[S(\epsilon)]$ determined by the (normalized) entropy $S(\epsilon)$, a function of energy $\epsilon$, as $\mathrm{d}\Gamma/\mathrm{d}\epsilon=G_G(S)\propto\exp[S(\epsilon)]/\Delta\epsilon$, holding under the ergodic assumption and valid for stochastic (Markovian) processes. 
The dependence of the phase space element on entropy $S(\epsilon)$ can be generalised, defining an arbitrary function $G(S)$ such that 
\begin{equation}\label{gamm}
\mathrm{d}\Gamma/\mathrm{d}\epsilon=G\big[S(\epsilon)-S(E)\big]/\Delta\epsilon,
\end{equation}
with $S(E)$ the entropy at any given reference energy $E$. Application to the Gibbs-function $G_{G}=\exp S$ just yields the arbitrary proportionality factor $\exp[-S(E)]$. 

The function $G(S)$ is subject to constraints which have to be determined from the requirement that the probability of finding a particle in energy state $\epsilon_i$ in phase space in the given fixed interval $\Delta\epsilon$ around $E$  is  
\begin{equation}\label{eq-prob}
w_i\propto \int\left[\frac{\mathrm{d}\Gamma(\epsilon)}{\mathrm{d}\epsilon}\right]\delta(\epsilon_i+\epsilon-E)\ \mathrm{d}\epsilon.
\end{equation}
Considering the product d$\Gamma_1(S_1)$ d$\Gamma_2(S_2)$ of two phase space elements of different entropies $S_1,S_2$ yields 
\begin{equation}
\mathrm{d}\Gamma_1\mathrm{d}\Gamma_2\propto G(S_1)G(S_2). 
\end{equation}
It is easy to prove that the only function for which this expression becomes equal to $\mathrm{d}\Gamma_3\propto G(S_3)$, with $G(S_3)=G(S_1+S_2)$, is Gibbs' $\log G_{G}(S)\propto S$. [This can be seen by assuming $S_2=S_1(1+\Delta/S_1)$ which produces an additional quadratic term in $G(S_1)G(S_2)$. For all $G\neq G_B$ this term is irreducible to a constant factor. An illustrational example is given below in Eq.(\ref{gamma3}).] For any $G(S)$ different from $G_{G}$ the entropies of the two phase space elements are not independent but correlated, indicating the presence of phase-space correlations, entropic non-extensivity, and violation of the ergodic assumption \citep[see, e.g.,][for a more extended discussion]{treumann2014}. 

Expanding the entropy around energy $E$ 
\begin{equation}\label{entexp}
S(\epsilon)\sim S(E)+\frac{\partial S(\epsilon)}{\partial\epsilon}\bigg|_{\epsilon=E}(\epsilon-E)=S(E)+\frac{\epsilon-E}{T},
\end{equation}
with $\partial S(\epsilon)/\partial\epsilon\big|_{\epsilon=E}=1/T$ and inserting into $G(S)$, Eq. (\ref{eq-prob}) yields the wanted phase space probability distribution  
\begin{equation}\label{eq-prob1}
w_i(\epsilon_i) \propto G(-\epsilon_i/T).
\end{equation}

For $w_i(\epsilon_i)$ being a \emph{real physical} probability,  the requirement is that $G(S)$ be a real valued positive definite function of its argument which can be normalised by summing over all accessible phase space energy states $\sum_iw_i(\epsilon_i)=1$. This determines the constant of proportionality in Eq. (\ref{eq-prob1}). Of the whole class of such functions $G$, only those may have \emph{physical relevance} which satisfy a number of supplementary constraints. 

The first {physical} constraint is that the probability should conserve energy. This is most easily checked for non-relativistic energies $\epsilon_i = p_i^2/2m$, where $m$ is particle mass. Under the simplifying ideal gas assumption [for non-ideal gases one just adds an external or an interaction potential field $\Phi(\mathbf{x})$ of spatial dependence] and, for $d$ momentum space dimensions, energy conservation implies that, for large $p$,  the function $G\{S[\epsilon(p)]\}$  \emph{asymptotically} converges faster than $p^{-d-2}$. 

Any  exponentially decaying asymptotic function of momentum $p$ or energy $\epsilon$ would thus be appropriate. In contrast to algebraic functions, for which more restrictive conditions would have to be imposed, they also hold under the requirement of conservation of higher moments of the probability distribution. One particular class of such functions is 
\begin{equation}
{G}(S)=\mathrm{e}^{S(\epsilon)}\tilde{G}(S),
\end{equation}
where $\tilde{G}(S)$ is any algebraic function. It produces the modified Gibbsian probability distribution $w_i\propto \tilde{G}(-\epsilon_i/T)\exp(-\epsilon_i/T)$. 

A severe restriction imposed on $G(S)$ is the demand that any stationary statistical mechanics based on Eq. (\ref{gamm}) is required to be {in accord with thermodynamics}. It needs reproducing the macroscopic relations between entropy $S$, energy $E$, pressure $P$, volume $V$ and the derived thermodynamic potentials. In addition, the temperature $T$ must maintain its thermodynamic meaning. 

The obvious path to statistical mechanics paralleling Gibbs' approach is via inverting $G(S)$, finding the appropriate entropy $S\left[w_i(G)\right]\propto S\left[G^{-1}(w_i)\right]$ as functional of  probability. This is suggested by Gibbs-Boltzmann's use of $\log w_{GB}$, which is the inversion of the Gibbs-Boltzmann probability $w_{GB}\propto \exp(-\epsilon/T)$, in the definition of entropy $S$. 

Thus, formally, for any arbitrarily chosen $G(S)$ that satisfies the above two constraints, the entropy is determined as the ensemble average 
\begin{equation}\label{eq-inv}
S=\bigg\langle G^{-1}\Big[\frac{w_i}{A}\Big]\bigg\rangle \propto \int \mathrm{d}\mathbf{p}\,\mathrm{d}\mathbf{x} \ w(\epsilon_\mathbf{p}) \,G^{-1}\left[\frac{w(\epsilon_\mathrm{p})}{A}\right],
\end{equation}
with $A$ the normalisation constant. It requires the existence of the inverse function $G^{-1}(S)$ which, for arbitrary $G(S)$, poses a  hurdle of constructing a viable statistical mechanics. Since $w(\epsilon_\mathbf{p})$ is normalised, a constant $-\log A$ can be added to $G^{-1}$ in order to adjust for the first law. Of interest is the entropy density 
\begin{equation}
s = w_i\,G^{-1}\big[w_i/A\big],
\end{equation}
rather than the total entropy $S$. This completely general definition of entropy may be of relevance not only for statistical mechanics but as well for several other fields like maximum entropy methods in data analysis, information theory, where it replaces Shannon's classical definition of information, and also outside physics in the economical and social sciences. Its extension to complex functions $G(S)$ implying complex probabilities is obvious. Since the meaning of complex entropies is unclear, it requires that the entropy $S$ in Eq. (\ref{eq-inv}) be real, i.e. calculated as a real expectation value. 

In the following we demonstrate that for a limited class of \emph{algebraic} functions, so-called generalised Lorentzians $G(x)= \left(1+x/\kappa\right)^{-\kappa}$, with arbitrary expression $x\equiv S$ independent of $0<\kappa\in\mathsf{R}$ \citep[cf., e.g.,][]{treumann1999a},  construction of a viable statistical mechanics is nevertheless possible if only the ubiquitous \emph{additional constraint} is imposed on $G(S)$ that, in some asymptotic limit, it reproduces the exponential dependence of Gibbs' phase space element on entropy. 

%
%

\section{Algebraic Example: Gibbs-Lorentzians}

This section is intended to demonstrate the above general theory to one particular algebraic case investigated in more detail in \citep{treumann2014}. We only show the particular form the function $G(S)$ assumes in this case.

With $G(S)$ an algebraic generalised Lorentzian, substitution into Eq. (\ref{gamm}) yields
\begin{equation}
\frac{\mathrm{d}\Gamma(\epsilon)}{\mathrm{d}\epsilon}=\frac{1}{\Delta\epsilon}\left\{1+\frac{1}{\kappa}\Big[S(E)-S(\epsilon)\Big]\right\}^{-\kappa-r}.
\end{equation}
It is obvious that for $\kappa\to\infty$ this expression reproduces the Gibbsian variation of the phase space volume. The negative sign in the brackets is self-explanatory. The relevance of entropy $S(E)$ is seen in that, for $\kappa\to\infty$, it just generates  a constant factor in d$\Gamma$. We also made use of the freedom of adding a number $r\in\textsf{R}$ to the exponent, as it has no effect when taking the large $\kappa$ limit. 

It is easy to prove explicitly that finite $\kappa<\infty$ imply correlations in phase space by (even for constant $\kappa$) considering 
\begin{eqnarray}\label{gamma3}
\mathrm{d}\Gamma_3&=&\bigg\{1+\frac{[S_3(E)-S_3(\epsilon)]}{\kappa} + \cr
&&+\frac{[S_1(E)-S_1(\epsilon)][S_2(E)-S_2(\epsilon)]}{\kappa^2}\bigg\}^{-\kappa-r}.
\end{eqnarray}
The irreducible quadratic term indicates that the two phase space elements in the $\kappa$-generalized Gibbs-Lorentzian model are not independent.

Eq. (\ref{eq-prob1}) yields
the Gibbs-Lorentzian probability  
\begin{equation}\label{gibbsdist}
w_{i\kappa}(\epsilon_i,\mathbf{x})=A\Big\{1+{\Big[\epsilon_i+\Phi(\mathbf{x}_i)\Big]}/{\kappa\ T}\Big\}^{-\kappa-r},
\end{equation}
generalised here to non-ideal gases by including a potential energy $\Phi(\mathbf{x}_i)$.  $A$ is a constant of normalization of the probability when integrating over phase space d$\Gamma$. Eq. (\ref{gibbsdist}) allows for a formulation of Gibbsian statistical mechanics consistent with fundamental thermodynamics. 

For the determination of the value of power $r=\frac{5}{2}$ one may consult \citep{treumann2014} or the Journal reference where the full paper was published.

These expressions also contain the \emph{Lorentzian power law} distributions, $w_{ir'}\propto [a+\epsilon_{i}/T]^{-r'}$ with $r'=\kappa+r$, encountered in cosmic ray spectra, for instance, where $a^{r'}=w_{ir'}(0)/A$. For $\kappa$ one obtains $r'=\frac{7}{2}$. Defining $\ell$ the number of the highest conserved statistical moment yields, more generally, $r'=1+\frac{1}{2}(2\ell+1)$.

Particular forms for $G(S)$ have been proposed in Tsallis' $q$ statistical mechanics \citep{tsallis1988, gell-mann2004} and in the Generalized Lorentzian thermodynamics \citep{treumann1999a}. Adopting the latter version we define the functional   
\begin{equation}
g[w]=\exp\left\{-\kappa\left[\bigg(\frac{A}{w_{\kappa}}\bigg)^{(\kappa+r)^{-1}}\!\!\!\!\!\!-1\right]-\log{A}\right\},
\end{equation}
whose logarithmic expectation value leads to the entropy $S=-\big\langle\!\log g[w]\big\rangle$.
Its particular version is chosen in agreement with Eq. (\ref{eq-inv}) for reconciling with thermodynamics by adding an additional normalization constant $A$. Clearly, $\log g$ is related to the inverse function $G^{-1}[w_\kappa/A]$ in this case. Substituting $w_{i\kappa}(\epsilon_i)$ and $g[w_{i\kappa}(\epsilon_i)]$ into the ensemble average Eq. (\ref{eq-inv}) yields
\begin{equation}
S = -\log {A} +{\langle E\rangle}/{T}.
\end{equation}
The thermodynamic relation $\langle E\rangle=TS+F$ identifies  $F =T\log A$ as the free energy $F$. The generalized canonical Gibbs $\kappa$-probability distribution then reads
\begin{equation}
w_{i\kappa}=\frac{\exp{(F/T)}}{\left(1+\epsilon_i/\kappa\ T\right)^{\kappa+r}}.
\end{equation}
Since $A$ is the normalization of $w_{i\kappa}$ one also has that $\sum_iw_{i\kappa}=1$ and, hence, for the free energy
\begin{equation}
F=-T\log \int \mathrm{d}\Gamma \bigg[1+\frac{\epsilon(\mathbf{p,x})}{\kappa\ T}\bigg]^{-\kappa-r} = -T\log Z_\kappa,
\end{equation}
with d$\Gamma=\mathrm{d}^3p\,\mathrm{d}V/(2\pi\hbar)^3$ the phase space volume element. From the last expression we immediately read the generalized Gibbsian version of the classical canonical partition function
\begin{equation}
Z_\kappa\equiv \int \mathrm{d}\Gamma \big[1+{\epsilon(\mathbf{p,x})}/{\kappa\ T}\big]^{-\kappa-r}.
\end{equation}
In the quantum case the integral becomes a sum over all quantum states $i$:
\begin{equation}\label{eq-z}
Z_\kappa\equiv\sum_i\big(1+{\epsilon_i}/{\kappa\ T}\big)^{-\kappa-r} .
\end{equation}
This completes the discussion for a system with fixed particle number since all statistical mechanical information is contained in the partition function $Z_\kappa$.

For further discussion see \citep{treumann2014} and the Journal reference where the full paper was published.

\section{Conclusions}
The important point here is that we provided a completely general recipe for constructing an equilibrium statistical mechanics from arbitrary functionals $G(S)$ with universal ``Gibbsian" phase space probability distribution Eq. (\ref{eq-prob1}). If only the inverse functional $G^{-1}(S)$ exists, the generalised entropy follows as its ensemble average (expectation value) allowing for the formulation of an equilibrium statistical mechanics. This form of entropy extends and generalises the Gibbs-Boltzmann-Shannon definition. It can be extended to complex functions $G(S)$ and complex probabilities under the requirement that the entropy obtained from Eq. (\ref{eq-inv}) is real. This version of entropy might be applicable not only in physics, but also in information theory, maximum entropy methods in data analysis, and possibly even in the economic and the social sciences.

As for an example we revisited the Gibbs-Lorentzian statistical mechanics \citep{treumann2008} which leads to $\kappa$-distributions of energetic particles. Such distributions result from wave particle interaction in plasmas \citep{hasegawa1985,yoon2012} and have been observed in the heliosphere \citep{christon1991,fisk2006}. They also apply to observed cosmic ray spectra.  

Gibbs-Lorentzian statistical mechanics is restricted to high temperatures only, excluding vanishing absolute temperatures. It thus categorically forbids any negative absolute temperatures $T<0$ as they would require cooling across the non-existing state $T=0$. Since $\kappa\to\infty$ reproduces classical and quantum statistical mechanics for all $T$, this conclusion provides another proof for the \emph{nonexistence of negative} absolute temperatures following from Gibbsian theory, supporting a recent proof \citep{dunkel2014} of this fact.
\paragraph*{Acknowledgement.}
This paper was part of a 2007 visiting scientist period of RT at ISSI, Bern. Hospitality of the ISSI staff is acknowledged.

\end{document}